\documentstyle[twocolumn,epsf,prl,aps]{revtex}

\begin{document}

\draft
\twocolumn[\hsize\textwidth\columnwidth\hsize\csname@twocolumnfalse%
\endcsname
\title{Stripe structure, spectral feature and soliton gap 
in high T$_c$ cuprates}
\author{Kazushige Machida, and
Masanori Ichioka}
\address{Department of Physics, Okayama University,
         Okayama 700-8530, Japan}
\date{\today}
\maketitle
\begin{abstract}
We show that the lightly doped 
La$_{2-x}$Sr$_{x}$CuO$_{4}$ can be described in terms of a stripe magnetic structure or 
soliton picture. The internal relationship between the recent  neutron observation of the diagonal ($x$=0.05) to 
vertical ($x\ge$ 0.06) stripe transition, which was predicted,
and the concomitant metal-insulator transition is clarified 
by this solitonic physics.
The phase diagram with the unidentified transition lines between antiferromagnetic to stripe phases, the doping dependence of the modulation period, the origin of the mid-infrared optical absorption are investigated comparatively with other single layer systems: La$_{2-x}$Sr$_{x}$NiO$_{4}$ and (La,Nd)$_{2-x}$Sr$_{x}$CuO$_{4}$.
The novel type of quasi-particles and holes is fully responsible for metallic conduction
and ultimately superconductivity.
\end{abstract}
]

Much attention has been focused on electronic properties in underdoped high T$_c$ cuprates recently because it is 
hoped that understanding of the unusual  ground state may lead us to a clue to a high T$_c$ mechanism. However, in spite of intensive theoretical and experimental works in the past little consensus for the 
normal state properties has emerged yet.

Recently a remarkable series of elastic neutron experiments~\cite{0.05,0.06,0.12} on La$_{2-x}$Sr$_{x}$CuO$_{4}$ (LSCO) has been performed, revealing static magnetic incommensurate (IC) structure: (1) For $x$=0.12, 0.10, 0.08 and 0.06, which are all metallic and superconducting, the modulation vector $\bf Q$ are characterized by ${\bf Q}=({1\over2},{1\over2}\pm\epsilon)$ and $({1\over2}\pm\epsilon,{1\over2})$  in reciprocal lattice units (r.l.u.). The IC modulation runs vertically in the $a$ ($b$)-axis of the CuO$_2$ plane. (2) For $x$=0.05 which is insulating the above four superspots rotate by 45$^{\circ}$ around the $({1\over2},{1\over2})$ position, 
characterized by $({1\over2}\pm\epsilon,{1\over2}\pm\epsilon)$. The IC 
modulation is now in the diagonal direction. (3) The incommesurability $\epsilon$ is given by  $\epsilon=x$
for $0.05\le x\le 0.12$ beyond which $\epsilon$ saturates 
(see Fig.2). 
(4) For  $x$=0.03 and 0.04 the elastic peaks are also observed at the commesurate (C) position, but the peak 
profiles are abnormally broaden, suggesting that
the IC modulation is hidden. The static nature of these orderings at low T's are also corroborated by $\mu$SR 
measurement~\cite{nieder}.

These lightly doped systems were known to be in spin-glass region ($0.02\le x\le 0.05$). In the further 
dilute region ($0< x\le 0.02$) upon lowering T the antiferromagnetic (AF) phase changes into a ``spin-freezing" structure at the 
Johnston line given by T$_f$=815$x$ (K), below which the internal magnetic field probed by the La-NQR is mysteriously broaden~\cite{johnston}. This transition line will be interpreted differently below.

Prior to these experiments, there has been known the 
static 
magnetic structures in superconducting 
(La, Nd)$_{2-x}$Sr$_{x}$CuO$_{4}$ (Nd) and insulating La$_{2-x}$Sr$_{x}$NiO$_{4}$ (Ni). In the former Nd system~\cite{tranquada1}
the vertical stripe with the same $\epsilon=x$ relation as in LSCO for $x>0.07$ and in the latter Ni system the diagonal stripe with $\epsilon=x/2$ up to $x\simeq 0.33$ 
are observed~\cite{tranquada}. Note that in YBCO$_{6.6}$ vertical IC fluctuations are also reported in inelastic neutron experiments~\cite{mook}.

Almost a decade ago one of the authors~\cite{machida,kato} predicts some of the above features of the static IC spin modulation 
in lightly doped cuprate systems, stressing the charged stripe or the solitonic structure as a convenient and universal ``vehicle" to accommodate excess carriers in an 
otherwise commesurate antiferromagnet within a mean-field treatment for a simple two-dimensional 
Hubbard model. Some features of them
were independently found by others~\cite{hfa}  at that time and are now confirmed by a more sophisticated method, such as DMRG~\cite{white}. 
Specifically we have predicted~\cite{machida,kato}: 
(1) The diagonal to vertical transition occurs 
as doping proceeds. 
(2) The incommesurability  $\epsilon$ is given by $\epsilon=n_h/2$ where $n_h$ is the hole 
density per site and proportional to $x$. 
In other words, once holes are 
introduced in CuO$_2$ plane the ground state is always described by a modulated IC structure. 
(3) The newly 
created solitonic midgap band situated between the 
insulator main gap accommodates excess carriers, keeping 
the AF formation energy intact. The band width of the midgap state for the vertical stripe is wider than for the diagonal stripe, implying that the former (latter) has a tendency toward a metal (insulator).

Encouraged by the success of our predictions, we  
now further extend it to a slightly more realistic case: 
namely, we introduce a next nearest neighbor hopping $t'$ to describe the metallic situation, which is certainly 
important to mimic the actual Fermi surface topology~\cite{ino}. (The previous solutions with only the nearest neighbor hopping $t$ 
were all insulating and fail to describe a metallic state~\cite{machida,kato,hfa}.)
Thus the main purpose of this paper is to 
establish the solitonic picture to resolve various 
mysteries  or paradoxes in lightly doped regions, mainly focusing on LSCO. These 
include (A) why the metal-insulator transition and the diagonal-to-vertical transition occur simultaneously at $x\simeq 0.05$, (B) why the spin freezing ($0<x<0.02$) below the Johnston line and the spin-glass regions exist, 
and (C) why the midgap state observed by
the mid-infrared optical absorption~\cite{uchida} grows progressively with $x$. Here we are going to give a self-consistent picture based on soliton physics,
which is quite different from an influential idea of frustrating charge segregation by Emery et al~\cite{emery}.

We start out with the standard Hubbard model $H=-\Sigma_{i,j,\sigma}t_{i,j}c^{\dagger}_{i,\sigma}c_{j,\sigma}+U\Sigma_{i}n_{i,\uparrow}n_{i,\downarrow}$ in two dimensions where $t_{i,j}=t$ for the nearest neighbor pairs $(i,j)$ and $t_{i,j}=t'$ for the next nearest neighbor 
pairs situated on a diagonal position in a square lattice. 
(The energy is measured by $t$ and $t'/t=-0.1$ and $U/t=3.0$ are chosen in the following.) The 
mean field $\langle n_{i,\sigma}\rangle=n_i+\sigma M_i$ at $i$-site is introduced where $n_i(M_i)$ is the charge 
(spin) density. Thus the one-body Hamiltonian $H_{MF}=-\Sigma_{i,j,\sigma}t_{i,j}c^{\dagger}_{i,\sigma}c_{j,\sigma}+U\Sigma_{i}(\langle n_{i,\uparrow}\rangle n_{i,\downarrow}+\langle n_{i,\downarrow}\rangle n_{i,\uparrow})$ is solved self-consistently in $k$-space under a given hole concentration or filling $1-n_h$ per site ($n_h$=0 is just the half-filling), assuming a spatially periodic structure for spin and charge modulations with $\epsilon=n_h/2$.

The resulting phase diagram in T vs $n_h$ is displayed 
in Fig.1 where the simple commensurate AF state denoted by C is never stable at T=0.
Upon raising T the vertical incommensurate (VIC)  or the diagonal incommensurate (DIC) phase are crossed-over to the AF because of 
an entropy effect. These spatially 
modulated spin structures or soliton structures are characterized by a periodically placed $\pi$-shifted domain wall (or a kink) which neatly accommodates  excess carriers. The charge carriers align perpendicularly 
to ${\bf Q}-({1\over2},{1\over2})$ (see Fig.1 in Ref. \cite{kato}). The transition from DIC to VIC upon increasing $n_h$ can be understood as arising from the competitive effects between the nesting of $\bf Q$-vector $({1\over2}\pm\epsilon,{1\over2}\pm\epsilon)$ in DIC at low doping and gapping at the $(1,0)$ saddle point position with  large density of states in VIC.
Note that the nesting deteriorates with $n_h$, necessarily inducing DIC to VIC transition.

The Johnston line T$_f=815x$ (K) for $0<x\le 0.02$ which 
indicates the anomalous broadening of the La resonance line~\cite{yoshimari} thus named a  spin freezing T$_f$, 
signals the change of the internal magnetic field distribution.
It may correspond to the C-IC transition line in 
Fig.1. Although the present numerical calculation gives it a first order, actual transition is of second order (see the detailed analytical discussions in Ref.\cite{machida}).
According to Fig.1 the known spin-glass region (0.02$\le x \le 0.04$)  is also described by the DIC structure. Indeed the recent neutron~\cite{0.05,0.06,0.12} and $\mu$SR~\cite{nieder} experiments
suggest this is the case as mentioned above.
Previous spin-glass identification simply reflects dirt effects.

The incommesurability $\epsilon$ is calculated as $\epsilon=n_h/2$ near half-filling at T=0.
The reduced $k$-space area spanned by this $\bf Q$-vector
is exactly filled by excess holes, differing 
from the simple so-called 2$k_F$-instability in one-dimensional 
case ($k_F$ the Fermi wave number).
It means that the charge stripe is completely filled with holes, never stabilizing the ``half-filled" stripe in our calculations~\cite{zaanen}. This 
is in sharp contrast with Tranquada's claim for Nd systems~\cite{tranquada1}. 
The observed $\epsilon$ is precisely 
given by $\epsilon=x$ both for LSCO and Nd systems at low doping levels of $x\lesssim 0.12$.
This implies 
 $\epsilon=n_h$ which corresponds to a half-filled 
stripe
if we identify $n_h=x$. The observation $\epsilon=x$ is also contrasted with the neutron result $\epsilon=x/2$ for 
Ni systems~\cite{tranquada}, which does agree with our estimate. This ``factor 2" paradox in LSCO and Nd is quite intriguing 
and may be important in evaluating the actual hole concentration in CuO$_2$ plane. Here we suggest a possible scenario to resolve it: Fig.2 shows the observed 
data of $\epsilon$ for three systems (filled symbols) and $n_{eff}$ (open ones) estimated by Uchida et al~\cite{uchida} through optical conductivity measurements. It is seen that according to 
Uchida's data under a given $x$ the supplied $n_{eff}$
in CuO$_2$ is doubled for LSCO and Nd compared to Ni at least for $x\le 0.1$. This immediately resolves the above paradox, 
namely, $n_h=2x$ for LSCO and Nd. The theoretical relation $\epsilon=n_h/2$ is valid for all three systems commonly and there is no half-filled stripe.
It is noted, interesting enough, from Fig.2 that a precise coincidence between $\epsilon$ (filled circles) and $n_{eff}$ (open circles) against $x$ is seen to be hold even beyond $x\sim 0.12$ up to $x\sim 0.25$ for LSCO and Nd. It means that beyond $x\sim 0.12$ doped holes do not come into the relevant magnetic band.

The spatial magnetic structure is characterized by a soliton form with many higher harmonics of the Fourier 
components
of the order parameters; $M_{lQ}$ (spin) and $n_{lQ}$ 
(charge). The period of the charge density is half that of the spin density.
The tending limits of those toward $n_h\rightarrow 0$ 
(half-filling limit) are given by $M_{(2l+1)Q}/ M_{Q}\rightarrow {1\over2l+1}$,  $M_{Q}/ M_{C}\rightarrow {2\over\pi}$ and $n_{2lQ}\rightarrow 0$ ($l=1,2,3, \cdots$)
where $M_{C}$ is the AF order parameter.
That is, the spin modulation becomes a squaring-up cnoidal wave form, far from a simple sinusoidal one and 
the charge modulation becomes irrelevant when $n_h\rightarrow 0$. These features well coincide with the previous analytic solution $sn(x,k)$ for $t'$=0~\cite{machida}.

Fig.3 shows the band structures for VIC and 
DIC.  The original single band is fold back into the small  reduced zone. In  the AF case at 
half-filling the large main gap separates it into the filled and empty bands. The 
doping creates a new band or the soliton band, seen
from Fig.3, in between the AF gap whose magnitude is relatively intact 
upon doping (also see Fig.7 in Ref. \cite {kato}). 
In DIC the midgap band is empty to accommodate holes and situated above the 
chemical potential $\mu(=0)$.
The midgap band width for VIC is substantially wider that that for DIC.  In VIC the dispersive midgap band touches 
$\mu$, making it metallic since the doping increases the soliton band width. 

In Fig.4 the one-particle spectral weight is shown on the energy  vs wave number plane along the symmetry lines for a metallic state:
The reorganized band opens a large gap at (0,1) and $(-{1\over 2},{1\over 2})$ positions where the solitonic midgap band also situates. The hold-backed bands get spectral weight toward ($-1$,1) direction after taking a plateau at (0,1). This characteristic feature here is strikingly similar to that observed by photoemission experiments~\cite{ino}.
The midgap band creates 
``quasi-particles" in its bottom at (0,1) and simultaneously the valence band also produces the same amount of ``quasi-holes" at the valence band top 
at $(-{1\over 2},{1\over 2})$. 
These newly created quasi-particles and -holes are mobile, fully responsible for metallic conduction and ultimately superconductivity. 
The Fermi surface pockets for these novel quasi-particles
and holes are situated around the $(0,1)$ and
$(-{1\over 2},{1\over 2})$ position respectively in the Brillouin zone. The spectral weight has a two-fold symmetric form. The midgap band does not touch $\mu(=0)$ at (1,0).

The aligned holes in VIC on a wall move easier than in DIC because of the difference of the effective hoppings (vertical vs diagonal hopping processes),
resulting in the different effective masses for two soliton bands. The band dispersion is very anisotropic; 
the effective mass perpendicular to the $\bf Q$-vector is much heavier than that parallel to the $\bf Q$-vector. The holes localized on a wall are mobile only along the 
domain wall, once metallicity is attained upon doping. These make VIC more metallic  while DIC insulating. In fact in our calculation in Fig.1 VIC (DIC) just corresponds to a metal (insulator), as coinciding with the observation\cite{0.05,0.06}. However, we remark that the exact one-to-one 
correspondence is not generic, but the tendency 
that VIC is easier to become metallic than DIC is generic, independent of the choice of $t'$.  Note that Ni systems are all DIC~\cite{tranquada} and insulating, and Nd for $x>0.07$~\cite{tranquada1} and YBCO$_{6.6}$~\cite{mook} are all
VIC and metallic, agreeing with this general rule.

In Fig.5 we show the optical conductivity for two dopings. As $n_h$ increases, the newly created midgap 
absorption from the valence (conduction)-to-soliton band transition process progressively grows inside the main AF gap at $\omega\sim1.7$. Its area corresponds to the hole concentration. 
These characteristics agree with the common optical observations for all three systems~\cite{uchida}. The long-standing mystery of the origin of the mid-infrared 
absorption can be attributed to the soliton band. 

In conclusion, we have found: 
(1) The metal-insulator transition (MIT) is accompanied by the profound magnetic structural change from the diagonal to vertical stripes as $x$ increases. Note, however,  that the actual MIT may be more complicated in the sense that 
the ${\bf Q}$ vector may change gradually, resulting in a second-order like MIT. It is interesting to carefully trace the superspots in neutron experiment for the critical concentration of $x$=0.05 under varying T. 
(2) The incommesurate spin modulation persists  down to $x\rightarrow 0$ at T=0, even deeply into the known ``commesurate" ($0< x\le 0.02$) and spin glass regions ($0.02< x\le 0.04$) under the assumption that the doped 
holes spread out the whole sample without any defects or impurities. 
We point out a possibility that the Johnston line may signal this commesurate-incommesurate transition. 
(3) The solitonic midgap band formed upon doping is responsible to the so-called mid-infrared absorption in optical conductivity 
measurements. 
(4) The incommesurability $\epsilon$; deviation from $({1\over 2},{1\over 2})$ position is
linear in the hole concentration $n_h$ per site at least 
in lightly doped region, coinciding with the recent neutron observation by Endoh and Yamada group~\cite{0.05,0.06,0.12}.

Finally we remark that judging from the optical data of the main insulating gap (Ni$\sim$4eV and LSCO$\sim$2eV) 
the Ni system may have larger $U$ than in LSCO~\cite{uchida}. This makes DIC in Ni stabler, keeping the system insulating up to higher doping level ($x>0.33$) (see Fig.5 in Ref.\cite{kato} where the critical concentration of the DIC-VIC transition is shown to increase with $x$). This may be one reason why heavily doped nickelates remain insulating. As for Nd system we predict a similar 45$^{\circ}$ rotation of superspots in 
neutron experiments and concomitant metal-insulator transition at lower dopings. If  these predictions are indeed confirmed experimentally, the present soliton picture must be a proper starting point to begin with in 
order to consider a high T$_c$ mechanism.

The authors thank many colleagues for enlightening discussions on the most recent updated experiments, including Y. Endoh, K. Yamada, S. Wakimoto, H. Kimura, 
K. Hirota, S. Uchida, A. Fujimori, A. Ino, J. Tranquada, M. Arai and M. Matsuda.


\begin{figure}
\begin{center}
\leavevmode
\epsfxsize=6.0cm  
\epsfbox{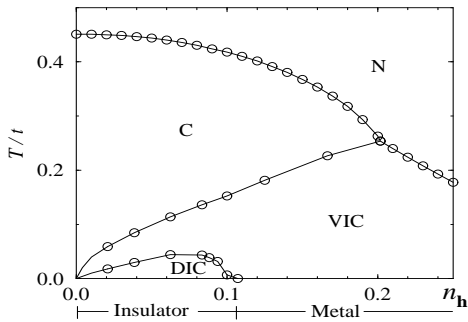}
\end{center}
\caption{
Phase diagram in T vs $n_h$. C: antiferromagnetic
VIC:vertical,  and DIC: diagonal incommensurate phases.
VIC (DIC) is metallic (insulating) at T=0. Note 
that $n_h=2x$ 
for LSCO.
}
\label{fig.1}
\end{figure}

\begin{figure}
\begin{center}
\leavevmode
\epsfxsize=6.5cm  
\epsfbox{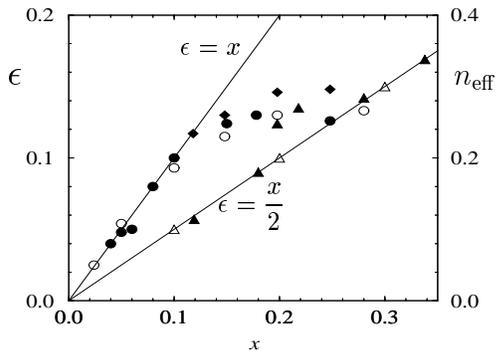}
\end{center}
\caption{
Incommensurability $\epsilon$ vs doping $x$ (left scale). 
$\epsilon=x$ for LSCO~\protect\cite{0.05,0.06} (filled circle) 
and Nd~\protect\cite{tranquada1} (diamond). $\epsilon=x/2$ for 
Ni~\protect\cite{tranquada} (filled triangle). 
The effective hole density $n_{eff}$ vs $x$ (right scale)
estimated by Uchida~\protect\cite{uchida} 
(open circle: LSCO, open triangle: Ni).
}
\label{fig.2}
\end{figure}

\begin{figure}
\begin{center}
\leavevmode
\epsfxsize=8.0cm  
\epsfbox{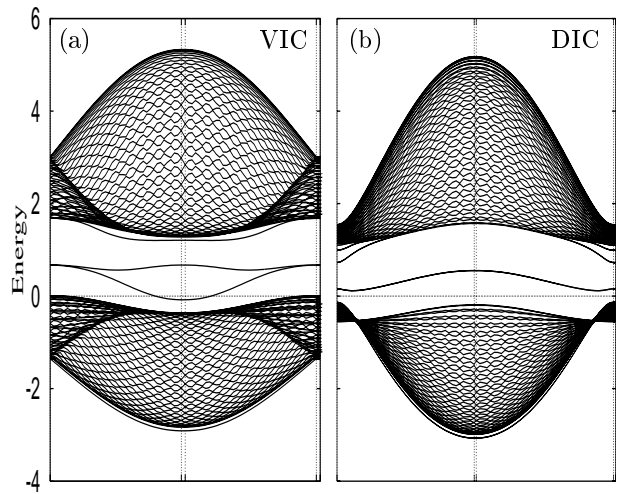}
\end{center}
\caption{
Dispersion relations for VIC(a) and DIC(b) 
along the perimeter of the reduced Brillouin zone for $n_h=0.02$. 
The chemical potential is at zero.
The midgap band is seen in between the AF gap.
}
\label{fig.3}
\end{figure}

\begin{figure}
\begin{center}
\leavevmode
\epsfxsize=8.0cm  
\epsfbox{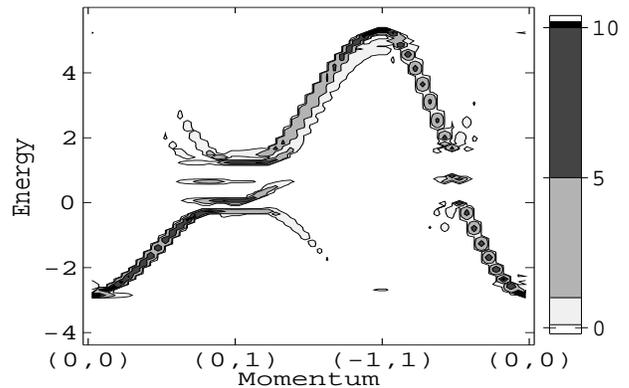}
\end{center}
\caption{
Contour map of the spectral weight along the symmetry lines 
in the original Brilloiun zone ($n_h=0.1$ and VIC).
The original cosine-like band is hold-backed and gapped to produce 
the midgap bands and the "shadow" bands
at around $\mu=0$.
}
\label{fig.4}
\end{figure}

\begin{figure}
\begin{center}
\leavevmode
\epsfxsize=6.0cm  
\epsfbox{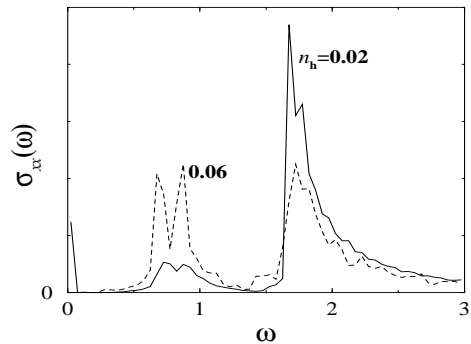}
\end{center}
\caption{
Optical conductivity (arbitrary units) vs frequency $\omega$
in VIC
for $n_h=0.02$ and 0.06. The midgap absorption at 
around $\omega\sim 0.8$ grows as doping proceeds.
}
\label{fig.5}
\end{figure}


\begin{thebibliography}{99}

\bibitem{0.05}S. Wakimoto, et al, preprint for 
$x=0.03,0.04,0.05$.

\bibitem{0.06}S. Wakimoto, et al, preprint for $x=0.06$.

\bibitem{0.12}T. Suzuki, et al, Phys.Rev.B{\bf 57},R3229 (1998) for $x=0.12$ and earlier references cited therein.

\bibitem{nieder}Ch. Niedermayer, et al, Phys.Rev.Lett.{\bf 80}, 43843 (1998).
They claim the similar phase diagram for YBCO.


\bibitem{johnston}J.H. Cho, et al, Phys.Rev.Lett.{\bf 
70},222 (1993) and F.C. Chou, et al, ibid {\bf 71},2323 (1993).


\bibitem{tranquada1}J.M.Tranquada, et al, Phys. Rev. Lett. 
{\bf 78},338 (1997) and earlier references cited therein.

\bibitem{tranquada}See reviews: J.M. Tranquada, cond-mat/9802043 and preprint (BNL-65377).

\bibitem{mook}H. Mook, et al, Nature {\bf 395},580 
(1998)
and cond-mat/9811100. M. Arai, private communication.

\bibitem{machida}K. Machida, Physica C{\bf 158},192 
(1989).

\bibitem{kato}M. Kato, et al, J.Phys.Soc.Jpn.{\bf 59}, 1047 (1990).

\bibitem{hfa}D. Poilblanc and T.M. Rice, Phys.Rev.B{\bf 
39},9749 (1989). J. Zaanen and O. Gunnarsson, ibid B{\bf 40},7391 (1990). H. Schulz, Phys.Rev.Lett.{\bf 64},1445 (1990).


\bibitem{white}S. White and D. Scalapino, Phys.Rev.Lett.{\bf 80},1272 (1998).

\bibitem{ino}A. Ino et al, preprint, who are just revealing
the detailed Fermi surface information in underdoped LSCO.


\bibitem{uchida}S. Uchida, et al, Phys.Rev.B{\bf 43},7942 (1991) for LSCO. T. Ido, et al, Phys.Rev.B{\bf 44},12094 (1991) for Ni. S. Uchida, private communication.

\bibitem{emery}See for example, V.J. Emery, et al, Phys.Rev.B{\bf 56},6120 (1997).

\bibitem{yoshimari}See B.J. Suh, et al, 
Phys.Rev.Lett.{\bf 81},2791 (1998)
for Li-doped LSCO.

\bibitem{zaanen}Also see, J. Zaanen and A.M. Ole\'{s}, Ann. Phys. {\bf 5}, 224 (1996).  T. Mizokawa and A. Fujimori, Phys.Rev.B{\bf 56}, 11920 (1997).

\bibitem{arima}T. Arima, et al, Phys.Rev.B{\bf 48},6597 (1993).

\end{thebibliography}
\end{document}